\documentclass[utf8]{frontiersSCNS} 
\usepackage{url,hyperref,lineno,microtype,subcaption}
\usepackage[onehalfspacing]{setspace}
\usepackage{longtable}
\usepackage{threeparttable}
\def\keyFont{\fontsize{8}{11}\helveticabold }
\def\firstAuthorLast{Y.-D. Hu et~al.} %use et al only if is more than 1 author
\def\Authors{Y.-D. Hu\,$^{1}$, E. Fern{\'a}ndez-Garc{\'\i}a$^{1}$, M. D. Caballero-García\,$^{1}$, I. Pérez-García\,$^{1}$, I. M. Carrasco-García\,$^{2}$, A. Castellón\,$^{2}$, C. Pérez del Pulgar\,$^{2}$, A. J. Reina Terol\,$^{2}$, and A. J. Castro-Tirado\,$^{1,2,*}$ (on behalf of a larger collaboration)}

\def\Address{$^{1}$Instituto de Astrof\'isica de Andaluc\'ia (IAA-CSIC), Glorieta de la Astronom\'ia s/n, E-18008, Granada, Spain \label{inst1} \\ %\email{huyoudong072@hotmail.com} \\
$^{2}$Unidad Asociada al CSIC Departamento de Ingenier\'ia de Sistemas y Autom\'atica, Escuela de Ingenier\'ias, Universidad de M\'alaga, C\/. Dr. Ortiz Ramos s\/n, 29071 M\'alaga, Spain  }

\def\corrAuthor{A.-J. Castro-Tirado}

\def\corrEmail{ajct@iaa.es}

\begin{document}
\onecolumn
\firstpage{1}

\title[BOOTES network]{The Burst Observer and Optical Transient Exploring System in the multi-messenger astronomy era} 

\author[\firstAuthorLast ]{\Authors} %This field will be automatically populated
\address{} %This field will be automatically populated
\correspondance{} %This field will be automatically populated

\extraAuth{}

\maketitle
\begin{abstract}

The Burst Observer and Optical Transient Exploring System (BOOTES) was first designed as an asset of autonomous telescopes that started to be deployed in 1998, taking 24 years to be fully developed around the Earth. Nowadays BOOTES has became a global network of robotic telescopes, being the first one present in all continents, as of 2022. Here we present the details of the network and review its achievements over the last two decades regarding follow-up observations of high-energy transient events. Moreover, considering the recent operations of neutrino and gravitational wave detectors, some hot-topic expectations related to robotic astronomy are discussed within the framework of multi-wavelength astrophysics.

\tiny
 \keyFont{ \section{Keywords:} Robotic Astronomy, Optical observations, BOOTES Network, Telescopes}
\end{abstract}

\section{Introduction}
The industrial revolution gave rise to technological advances, leading to the coinage of the word 'robot' for the machines to replace workers in repeatable roles \citep{1992VA.....35..399B}. In science, robots quickly became popular, and one of the very first ones was a mobile robotic chemistry machine at Liverpool University which was designed to undertake repeatable chemical experiments in order to saving time and avoiding operating errors \citep{2020Natur.583..237B}. In Astronomy, attempts to achieve some degree of automation were undertaken since the mid-20th century, especially regarding space satellites which can be treated as robotic systems because they can operate with self-power supply, command uploading and remote control~\citep{2003A&G....44c..23E}. These robotic systems can augment or replace human activity in space, such as the robotic arm on the international space station (ISS) which can install and replace equipment and perform external inspections of the station. In addition to their use in space, systems for ground-based telescopes have been developed to allow certain tasks to be executed automatically. Thus, four degrees of automation have been achieved over the last decades, ranging from automatic tasks and automated telescopes to remote instruments and robotic observatories, depending on their degree of automation and the level of human interaction \citep{1992VA.....35..399B,2010AdAst2010E..60C}. Since the 1980s, there have been several old telescopes that could be upgraded and operated robotically. The advancement of robotic telescopes became possible with the development of internet and computer technology, leading to the design and construction of several ground-based robotic telescopes, such as the BOOTES network discussed here.

\section{The BOOTES network}
The Burst Optical Observer and Transient Exploring System (BOOTES) network \citep{1996AIPC..384..814C,2012ASInC...7..313C} is a worldwide robotic telescope network whose first prototype was proposed and designed under a Spanish-Czech collaboration framework. It achieved its first light in 1998. As the Spanish pioneer robotic observatory for optical transient searching and follow-ups, it has achieved multiple scientific goals, as detailed below.

\begin{figure}[!h]
	\includegraphics[width=\textwidth, clip]{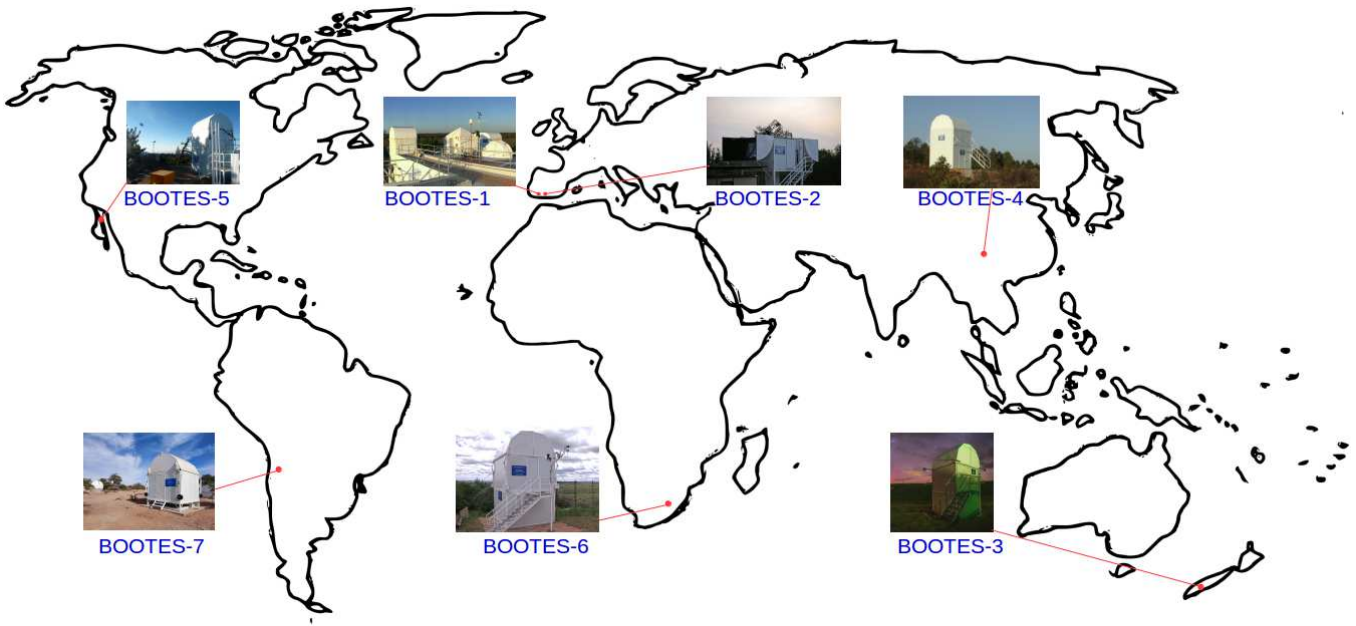}
	\caption{The BOOTES network map at present (2022), where all BOOTES network astronomical stations worldwide are marked with red points. }
	\label{fig:BOOTESmap}
\end{figure}

As originally planned, the BOOTES network consists of seven stations eventually, four in the northern hemisphere and three in the southern hemisphere ensuring that there will always be at least one telescope covering the northern and southern parts of the night sky \citep{2012ASInC...7..313C}. All stations are marked in Figure~\ref{fig:BOOTESmap}. Since the first light of BOOTES-1 in 1998 until the installation of BOOTES-7 in 2022, the BOOTES network has already deployed all its seven astronomical stations. These observatories are detailed in the following Table.

\begin{table*}
	\caption{BOOTES network sites location.}
	\centering
	\begin{tabular}{cccccc}
		
		\hline
		Site& latitude & longitude &Plus Codes& ASL  & Site\\
		&&&&(m)&\\
		\hline
		BOOTES-1&$37^{\circ}$05'58.2''N&6$^{\circ}$44'14.89''W&8C9M37X8+2C&50&Mazag\'on\\
		BOOTES-2&36$^{\circ}$45'24.84''N&4$^{\circ}$02'33.83''W&8C8QQX44+GQ&70&Algarrobo-Costa\\
		BOOTES-3&45$^{\circ}$02'22.92"S&169$^{\circ}$41'0.6''E&4V6FXM6M+QP&360&Lauder\\
		BOOTES-4&26$^{\circ}$41'42.8''N&100$^{\circ}$01'48.24''E &7PR2M2WJ+FF&3200&Lijiang\\
		BOOTES-5&31$^{\circ}$02'39''N&115$^{\circ}$27'49''W&85362GVP+JG&2860&Baja California\\
		BOOTES-6&29$^{\circ}$02'20''S&26$^{\circ}$24'13''E&5G28XC63+FF&1383&Maselespoort\\
  		BOOTES-7&22$^{\circ}$57'09.8''S&68$^{\circ}$10'48.7''W&2RWC+V7X&2440&Atacama\\
		
		\hline
		\multicolumn{4}{c}{}\\[4pt]
		
	\end{tabular}
	
	\label{tab:BOOTESsites}		
\end{table*}

\begin{figure}
	\centering
	\includegraphics[angle=0,scale=0.3]{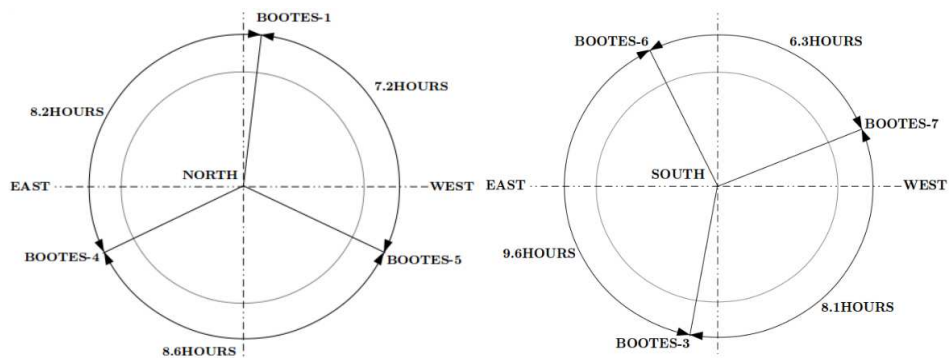}
	\caption{The location of BOOTES sites in the northern hemisphere (B1, B4, B5) and the southern hemisphere (B3, B6, B7) ensures that at least one telescope can monitor both the northern and southern skies at any night time (monsoon permitting) \citep{2014RMxAC..45...87H}.}
	\label{fig:BOOTESsites}
\end{figure}

\subsection{The BOOTES-network construction}
\subsubsection{The seven BOOTES astronomical stations}

The BOOTES-1 observatory (B1) is located at Estaci\'on de Sondeos Atmosf\'ericos in Centro de Expermentaci\'on de El Arenosillo which belongs to Instituto Nacional de T\'ecnica Aerospacial (CEDEA-INTA) in Mazag\'on, Huelva, Spain \citep{2016AdAst2016E..12J}. It contains three domes not far from each other, BOOTES-1A (B1A), BOOTES-1B (B1B) and BOOTES-1C (B1C). B1A is equipped with two wide-field CCD cameras (4096$\times$4096 pixels$^{2}$) in the same mount, one attached to a 400 mm f/2.8 lens which covers a 5$^{\circ}\times$5$^{\circ}$ field of view (FOV) and another connected with a 135\,mm f/2 lens covering a 15$^{\circ}\times$15$^{\circ}$ FOV. In the B1B dome, there is a 0.3 m diameter Schmidt-Cassegrain reflector telescope mounted on a Paramount mount which covers a 15'$\times$15' FOV. Both cameras are working with a clear filter which can be transformed to the R-band under the assumption of no intrinsic colour evolution of the optical counterpart. B1C is the dome which was used to store the Spanish-Polish collaboration project 'Pi of the sky' which was a system of robotic telescopes with a wide field of view containing 4 units (16 CCD cameras) since 2010. After their retirement in 2020, this dome was re-furnished with a new mount pier and now hosts a 28 cm wide field of view camera operating since mid-2022.

\begin{figure}
	\centering
	\includegraphics[angle=270,scale=0.25]{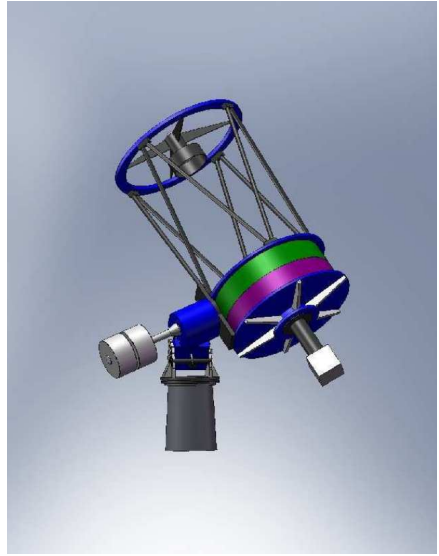}
	\caption{The ultra-light weight telescope concept was first used at the B2 station being the station we replicated in the rest of the BOOTES stations worldwide \citep{2012ASInC...7..313C}.}
	\label{fig:BOOTESmodel}
\end{figure}

The BOOTES-2/TELMA observatory (B2) is located at the Instituto de Hortofruticultura Subtropical y Mediterr\'anea La Mayora, which belongs to the Consejo Superior de Investigaciones Cient\'ificas and Universidad de M\'alaga (IHSM/UMA-CSIC) in Algarrobo-Costa (M\'alaga, Spain) which started its scientific operation in 2002~\citep{2016AdAst2016E..12J}. A 0.3\,m diameter Schmidt-Cassegrain telescope was first deployed with an attached wide-field camera similar to the one installed at B1B. The idea was to get B2 operating in two different modes: the stand-alone observation and the parallel stereoscopic mode. The latter mode allows for simultaneous observation together with the B1B telescope, which is located at a distance of 250 km. This setup allows for the discrimination of near-Earth detected objects up to a distance of $10^{6}$ km. In 2008, a new high-speed slewing fast-camera and fast-filters telescope prototype was purchased in order to upgrade the 0.3\,m telescope to quickly follow-up astronomical transients. The new telescope installed was a 0.6\,m aperture Ritchey-Chretien one with f/8 and its optical tube truss made of carbon fibre making it a lightweight instrument with an overall weight of about 70 kg (see Figure~\ref{fig:BOOTESmodel}). The equatorial mount NTM-500 from the Astelco company was chosen because it had the ability to achieve speeds up to 30 deg/s and accelerations up to 10 deg/s$^{2}$ according to the manufacturer. Its pointing accuracy is less than 5", and the tracking accuracy is less than 1" per hour once a proper pointing model is achieved. A wide temperature range of -20$^{\circ}$C to 40$^{\circ}$C is suitable for its operation. Meanwhile, the Andor iXon X3 EMCCD 888 was attached to the telescope in order to capture images on the 1024$\times$1024 pixels CCD detector, which has a pixel size of 13-$\mu$m and a full resolution frame rate of 9 fps, thus providing a FOV of 10'$\times$10'. With such an ultra-light telescope, the mount has the capability to achieve fast slewing speeds and accelerations to reach any part of the sky in less than 8 seconds.

The BOOTES-3/YA (B3) observatory was the first observatory of the BOOTES network in the southern hemisphere. It was deployed in February 2009 at Vintage Lane, Blenheim (New Zealand \citealt[27 m.a.s.l.;][]{2012ASInC...7...79T}). Due to unavoidable environmental factors, it was moved from the Northern part of the South Island to the Southwest side of the South Island, with the site lying at the National Institute of Water and Atmospheric Research (NIWA) in Lauder, nearby Otago, since 2014. The same telescope and mount as that at BOOTES-2 were chosen thus making B2 the continuing prototype for the BOOTES network since then (see Figure~\ref{fig:BOOTESmodel}). Hence, BOOTES-3 contains a 0.6\,m aperture, f/8 beam Ritchey–Chretien telescope atop an Astelco NTM-500 mount. A variety of filters, from clear to SDSS u' g' r' i' to WFCAM/VISTA Z and Y can be attached to an Andor 1024$\times$1024 pixel$^{2}$ CCD camera, thus able to provide multi-wavelength photometric observations within a 10'$\times$10' FOV.

The BOOTES-4/MET (B4) observatory is located at the Lijiang Astronomical Observatory in Lijiang, China and is operated by the Chinese Academy of Sciences. It was the second observatory of its kind in the northern hemisphere outside of Spain, and the first Chinese robotic astronomical observatory. It has been in operation since February 2012. This astronomical site is located approximately 120 degrees east of the BOOTES-1 observatory, resulting in a time difference of around 8 hours. In addition to this, there is also a 0.6-meter Ritchey-Chretien telescope with an Andor camera mounted on an Astelco mount at the Cassegrain focus of the telescope.
Similar to B2 and B3, the FOV of this telescope is 10'$\times$10' with a filter set comprising a clear filter plus SDSS filter set (u' g' r' i') and both WFCAM/VISTA Z and Y filters.

The BOOTES-5/JGT (B5) observatory is the third observatory in the northern hemisphere and is located at the National Astronomical Observatory in Sierra de San Pedro M\'artir (Bajo Califomia, M\'exico \citealt{2014RMxAC..45...87H,2016RMxAC..48..114H}). This astronomical site was chosen due to its longitude, which is approximately 120$^{\circ}$ west from the B1 observatory. It began operations in Nov 2015 and is equipped with the same equipment as the BOOTES-3 and BOOTES-4 stations. As shown in Figure \ref{fig:BOOTESsites}, the three BOOTES stations in the northern hemisphere (B1, B4, B5) are located at roughly the same latitude with evenly divided longitudes. This placement secures that the BOOTES network can monitor the northern sky at all times with at least one observatory always available.

The BOOTES-6 (B6) observatory was the second astronomical station in the southern hemisphere and is located at Boyden Observatory, Maselspoort (Bloemfontein, South Africa). With a similar setup to B2, B3, B4 and B5 stations, it was deployed in late 2021.

Finally, the BOOTES-7 (B7) observatory is located in San Pedro de Atacama (Chile) and was deployed in late 2022. It is equipped with the same telescope, camera, and filter set as the other observatories of the network in order to ensure complete coverage of the southern sky, just as the network of observatories in the northern hemisphere already do, as shown in Figure \ref{fig:BOOTESsites}. 

The information for all BOOTES network sites is listed in Table \ref{tab:BOOTESsites}, and the information for the telescopes is listed in Table \ref{tab:telescopeparameter}.

\subsubsection{The enclosures}
The enclosures at the BOOTES observatories consist of two halves with an overlapping roof that can open or close in a shell fashion under the action of electric motors. The overlapping direction of the two halves is set to the upwind direction of the site's prevailing wind. Additionally, the two halves are designed to open fully, allowing the telescope to access any part of the sky with an airmass of less than 5.8. The two halves of the enclosures at the BOOTES observatories each have two electric motors connected through hinges and gears that provide the necessary torque to start and finish observations. The B2 dome, however, uses two oil pressure pumps and extra mechanical construction instead of motors. These motors/pumps are controlled automatically by the system, but they can also be activated manually during upgrading and commissioning inside. At each station, a weather station is mounted on a metallic tower near the dome position as an important part of the security system (see the description in the section below). The weather station includes a meteorological camera and precipitation and cloud sensors that work together to determine wind, cloud cover, rain, and humidity conditions. Besides, the outside dome surveillance camera for checking the dome's situation is also mounted at the same place. Another surveillance camera is installed inside the dome. The measured parameters from both units constitute the selection criteria for the control system to open/close the dome in less than 30 s.

\subsubsection{The CASANDRA very wide-field cameras}
In addition to a 0.6 m telescope, each BOOTES station is equipped with an all-sky camera named Compact All-Sky Automated Network Developed for Research in Astronomy \citep[CASANDRA,][]{2011AcPol..51b..16C}. A 4096$\times$4096 pixel$^{2}$ CCD camera attached to a 16\,mm f/2.8 lens provides a 180$^{\circ}$ FOV which can detect bright stars with a magnitude \textless8 mag near the horizon and approximately 10 mag at the zenith. The working mode of this camera is programmed to take images every minute which are used to monitor sky conditions and detect astronomical events, such as meteors.

\begin{figure}
	\centering
	\includegraphics[angle=0,scale=0.3]{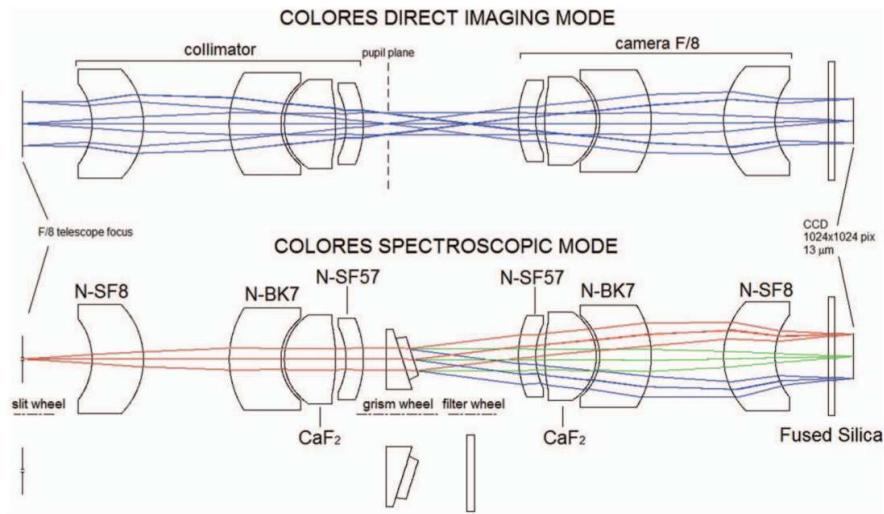}
	\caption{The optical path configuration of COLORES in direct imaging and spectroscopic modes \citep{2013RScI...84k4501R}.}
	\label{fig:coloresopticalpath}
\end{figure}

\subsubsection{The COLORES imaging spectrograph}
Spectroscopic observations can offer crucial insights into the nature of the sources, but fast follow-up observations are hindered by the time-consuming process of changing instruments and other factors on medium/large telescopes. This can create significant difficulties in obtaining spectroscopic observations of rapidly fading transient events during their early stages. Consequently, spectroscopic instruments mounted on robotic telescopes benefit from quicker reaction times and more efficient scheduling, which enhances the ability to detect fast-evolving transients. In this context, we want to highlight the COmpact LOw REsolution Spectrograph \citep[COLORES][]{2013RScI...84k4501R,2014PhDT.......140J,2014arXiv1408.4370C} we developed and mounted on the BOOTES-2 station. COLORES is a prototype for a light-weight spectrograph mounted on a small-size telescope and it has been functioning well so far. It is a Faint Object Spectrograph and Camera (FOSC) with a single optical path and wheels at multi-position installed. The slits, filters and grisms are inlaid in the wheels into the collimated space to prevent any changes in the focus plane. The idea is to make the instrument work in both imaging and spectroscopy modes, with the central object in imaging mode being the same as the object whose spectrum is observed in spectroscopic mode. As shown in Figure \ref{fig:coloresopticalpath}, three designed wheels are staggered vertically to the parallel beam, i.e. aperture, grism and filter wheels. Each one has 8 holes with a size of 2.54 cm$\times$ 2.54 cm per item, which can be replaced without dismounting the entire wheel. The aperture wheel is placed in the front of the collimator which is located at the Cassegrain focal plane. Here, 5 slots are occupied by the slits with widths of 25 $\mu$m, 50 $\mu$m, 75 $\mu$m and 100 $\mu$m and length of 9.3 arcmins (to suit the different atmospheric seeing conditions) and also a blank plate. Meanwhile, one empty slot is used for imaging mode and the other two slots for a future equipment upgrade. The grism wheel follows the collimator and has four grisms currently available, i.e. GTK19, P-SF68, SF2 and N-BK7, to cover the wavelength range of 3800-11500\,\AA \, with a spectral resolution of 15-60\,\AA \ under the combination of a fixed slit and one grism. The filter wheel is in between the grism wheel and the camera which has 7 slots used for clear, SDSS g' r' i', Bessel R, WFCAM/VISTA Z and Y filter together with an empty place for the light to pass in spectroscopic mode. During the direct imaging mode and acquisition image mode, the aperture wheel and the grism wheel remain fixed with their empty slots. When the telescope switches to the spectroscopic mode, the filter wheel is set to an empty slot and the chosen slit and grism are engaged. The wavelength calibration is achieved through calibrating the standard lamp spectrum. In front of the aperture wheel, two tubes in COLORES have been placed opposite each other and oriented perpendicular to the light path. There are two standard lamps, i.e. Krypton and HgAr lamps, installed in one of the tubes controlled by the electronic program. Another tube contains a 45$^{\circ}$-tilted flat mirror which can be inserted into the optical path to reflect the standard lamp light to the camera or removed from the main light path by means of a motorized precision slide.

The COLORES imaging spectrograph was installed in B2 in 2012~\citep{2014PhDT.......140J}. This low-resolution spectrograph is capable of providing preliminary estimates of the distance to bright cosmological gamma-ray bursts and rough estimates of the chemical abundance information with a resolution of 15-60 \AA. For the time being, COLORES is only installed in B2, but in the future there will be a similar model installed in one of the Southern Hemisphere BOOTES observatories. Consequently, the BOOTES network with such light-weight imaging spectrographs shall provide spectroscopic observations for fast transients located at any position in the sky. Note that all the information about the telescopes mentioned in this section is listed in Table \ref{tab:telescopeparameter}.
\begin{table*}[h]
\begin{threeparttable}[b]
	\caption{Features of the BOOTES network of telescopes.}
	\centering
	\begin{tabular}{ccccccc}
		
		\hline
		Site& B1A & B1B &B2& B3/4/5/6/7 & CASANDRA\\
		%&&&&&&\\
		\hline
		Type\tnote{a}&Ph&SC&RC&RC&Ph\\
		Lens&400mm+135mm&-&-&-&16mm\\
		Mirror&-&30cm&60cm&60cm&-\\
		Focus\tnote{b}&-&Ca&Ca&Ca&-\\
		Focal ratio &f/2.8+f/2&f/10&f/8&f/8&f/2.8\\
		CCD&4096x4096&512x512&1024x1024&1024x1024&4096x4096\\
		Pixel size &9$\mu$m&16$\mu$m&13$\mu$m&13$\mu$m&9$\mu$m\\
		Angular Resolution&4.39"+13.2"&2"&0.59"&0.59"&2.2'\\
		FOV&5$^{\circ}$x5$^{\circ}$+15$^{\circ}$x15$^{\circ}$&17'x17'&10'x10'&10'x10'&180$^{\circ}$\\
		Filter\tnote{c} &C&C&g'r'i' C& u'g'r'i'&C\\
        &&& R ZY &C ZY&\\
		Mount&Paramount&Paramount&Astelco&Astelco&-\\
		&ME&MX+&NTM-500&NTM-500&\\
		Camera\tnote{d}&MG4& A887 &A888&A888&MG4\\
		Spectrograph &-&-&COLORES&-&-\\
		\hline
		%\multicolumn{4}{c}{}\\[4pt]
	\end{tabular}
	\label{tab:telescopeparameter}	
	\begin{tablenotes}
       \item [a] Photolens (Ph), Schmidt-Cassegrain (SC), Ritchey-Chretien (RC). %Nasmyth–Cassegrain (NC).
       \item [b] Cassegrain focus (Ca). %Nasmyth focus (Na),
       \item [c] Clear (C).
       \item [d] Moravian G4-16000 (MG4), Andor iXon EMCCD DV887(A887), Andor iXon X3 EMCCD 888 (A888).
     \end{tablenotes} 
	\end{threeparttable}
\end{table*}

\section{The BOOTES control system}
The control system is an important part of a robotic telescope network, responsible for ensuring that all the telescopes are operating correctly. When BOOTES was first established, the control system was known as the optical transient monitor~\cite[OTM,][]{2001grba.conf..412P} which was used for the first two wide-field prototype telescopes. Then, a Linux-based platform called Remote Telescope System - 2nd version~\citep[RTS2,][]{2006NCimB.121.1501K}, was used to control all telescopes with the instrument driver programmed in C++ language. Currently, the BOOTES network has been upgraded to a user-friendly, ASCOM platform-based system that contains an interface for communicating with the telescope server host on the Windows operating system. Every BOOTES station has a copy of this system installed on its host server.

This system is composed of three sub-systems: 1) Targets manager 2) Object executor and 3) Dome controller. 

The targets manager is running on the BOOTES site host server which runs a program to receive alerts automatically and supports a web-browser interface for communication with telescope users. 
Mainly, the targets manager is continuously listening to the output from the Gamma-ray Coordinates Network (GCN)/ Transient Astronomy Network (TAN) \footnote{https://gcn.gsfc.nasa.gov/} which distributes the locations of transients detected by various spacecraft (\textit{Swift, Fermi, MAXI, INTEGRAL, IPN,} etc) and ground-based multi-messenger detectors (\textit{LIGO/Virgo, IceCube, HAWC,} etc). Once a new transient position is received, the DakotaVoEvent module reacts to this alert by making it a target of opportunity (ToO) and assigning it the highest priority in the observation if its detectability and time window are suitable for the current site. By adding this higher priority target to the pending list, the ongoing observation will be aborted and the telescope will point to the ToO event. If there are not any new ToO objects, the telescope will be running with a prioritized observation list. As a backup, the BOOTES network has also installed a central server which communicates with all sites by sending any non-ToO and non-scheduled but interesting targets (with specific priority values) to a suitable observatory site based on their detectability. This method is generally used during the follow-up of objects within a large field.

The second part of the control system is the object executor, which is a program scripted in VisualBasic language, that drives the telescope mount and calls the Maxlm DL module to operate the CCD camera with a specific filter. The focuser module, FocusMax, is also steered with this software during the observation. Meanwhile, the calibrated universal time and geographic information are also provided by the GPS module. All of these communications are carried out using the common object model standard through the application programming interface. 

\begin{figure}[!h]
	\includegraphics[width=\textwidth, clip]{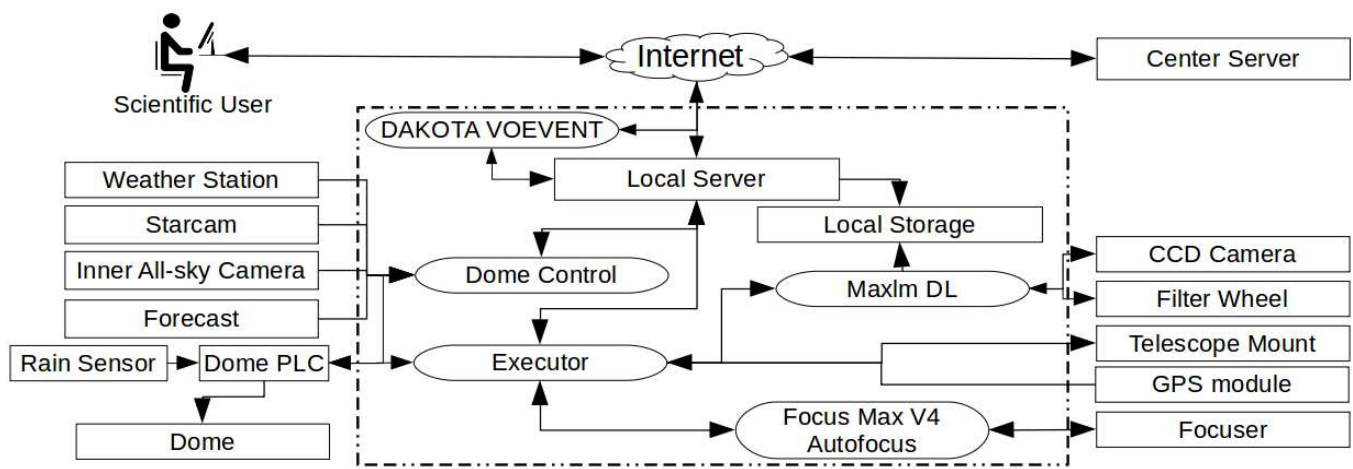}
	\caption{The control system is sketched in each observatory site. The dashed square contains the host server, which runs all modules of the telescope. The arrows mark the direction of the data flow and/or control commands.}
	\label{fig:BOOTEScontrolsystem}
\end{figure}

The dome controller is responsible for opening and closing the dome depending on environmental conditions. It plays a crucial role as others in this system to protect the telescope from bad weather, such as storms, and ensures that observations can be performed safely. This part of the system gathers information (such as cloud rate, temperature, wind speed and humidity) from the weather station, the outer all-sky camera, the inner all-sky camera and a forecast report from the internet. It then sends this information to the database to determine if the weather conditions are suitable for opening the dome and conducting observations. The database provides quick and frequent feedback to close the dome in case the weather conditions will get worst. As a final safeguard, a rain sensor is also included which takes priority to close the dome in the event that the other weather monitoring systems fails to detect rain. 
{\bf During the observation, the MaxmlDL program gathered data and saved it as a standard .fit image and the DS9 software transferred it to a companion .jpg format file as a snapshot, both of which are generated simultaneously and are kept in the host server. After the observation ends, the authorized scientific user can view the snapshot and/or download the .fit files immediately via the user-specific link located on the main page \footnote{http://bX.bootestelescopes.net/ (With "X" ranging from 1a to 7 depending on the given BOOTES telescope.)} of the HTTP server at each station.} Briefly, a sketch is made as shown in Figure~\ref{fig:BOOTEScontrolsystem} to depict the process operating within the control system described in this section. In the sketch, the rectangle represents the hardware and the ellipse represents the software involved. Meanwhile, the arrow lines indicate the direction of the data flow and the instructions set.

Normally, there are two working modes of initialisation in the BOOTES system: the ToO mode, which has a higher priority, and the monitor mode. The ToO mode is activated when a new alert is received on the host server and the top priority target is assigned. If the object's position is attainable by the telescope at the time, the ToO mode will interrupt the current observational plan. Otherwise, this object will be moved into the pending monitor list until the next observation time window comes. The monitor mode only operates in the non-ToO periods. In this mode, the telescope executes the planned objects in order of their priorities from the scientific community. Depending on the triggering instrument, the received coordinates of the new event from GCN have an error region ranging of a couple of arcminutes (such as BAT/Swift) to several degrees (such as LIGO/Virgo). This error region can be enhanced by the object identification through other follow-up observations. For example, the X-ray observations of gamma-ray bursts can locate the event within several arcseconds. For the new ToO observation, the BOOTES network takes images to cover the entire error region. If there is not any counterpart confirmed by other facilities, the obtained images from the BOOTES network are used to find the new object through the difference imaging method, where a late-time image or resampling images from the Pan-STARRS and the 2MASS catalogs within the same field of view will be used to calculate the residual image. During the monitor mode operations, the difference imaging method can also be used to detect any kind of optical flashes in the image. Once the optical counterpart is identified, the telescope points to the updated position and continues monitoring it. As a consequence of being a global network, the time difference between different sites in the northern and southern hemispheres allows for long-term monitoring of any kind of object.

\section{Scientific goals and results of the BOOTES network}
With the capability of fast slewing and quick reaction, the scientific goals of the global BOOTES network were first set as below:

a) Observations of GRB optical counterparts: from the prompt emission to the afterglow. Simultaneous multi-wavelength detections of optical counterparts to GRBs have been obtained in some cases with white band magnitudes in the range of $\sim$5--10. These observations provide important results on the central engine of the violent emitters which can be executed and monitored by BOOTES telescopes due to their fast slewing capability.

b) The detection of optical transients of astrophysical origin. These events could be related to new astrophysical phenomena, perhaps associated with Fast Radio Burst (FRB), Neutrino sources, Gravitational waves (GWs), Quasars (QSOs), Active Galactic Nuclei (AGN) and Tidal Disruption Events (TDEs). 

c) Ground-based support for the space missions, including the ESA's  
International Gamma-Ray Laboratory ({\it INTEGRAL}) and the NASA's Neil Gehrels Swift Observatory ({\it Swift}) satellites and also future missions, such as the Space Variable Objects Monitor ({\it SVOM}), to monitor high energy sources. 

d) Monitoring astronomical galactic objects which include meteors, asteroids, comets, variable stars and novae, etc.

Since its first operation in 1998, there have been fruitful scientific results and outreach projects achieved during its 24 years of existence. Here the main topics in the field of astrophysical transients are listed in the following sections, as well as the BOOTES network's contributions to public outreach.

\subsection{Gamma-ray bursts}
Gamma-ray bursts (GRBs) are the most energetic explosion phenomena in the Universe which have a duration from several seconds even up to thousands of seconds in their gamma-ray emission~\citep{1973ApJ...182L..85K,2009ARA&A..47..567G}. Normally, they can be classified into two categories~\citep{1993ApJ...413L.101K}: short GRBs \citep[SGRB, \textless 2s;][]{2014ARA&A..52...43B} and long GRBs \citep[LGRB, \textgreater2s;][]{2006ARA&A..44..507W} based on their temporal scale. Their gamma-ray emission is followed by a longer-lived fading emission (detected from the X-ray to the radio domain) which is called the 'afterglow'. This afterglow is produced when the relativistic fireball interacts with the surrounding medium to generate external shocks which can be used to pinpoint and study the GRBs and their host galaxies' properties~\citep{1999PhR...314..575P}. Furthermore the afterglow can be detected even months after the burst. Nowadays, the LGRBs have bright afterglows that can be studied in great detail. It has been found that LGRBs are associated with Type Ib/c supernova explosions which indicate that the progenitors are collapsed massive stars~\citep{2006ARA&A..44..507W}. SGRBs have been linked with the merger of compact objects, such as neutron star binaries (BNS) or neutron star-black hole binaries (NS-BS)~\citep{1992ApJ...395L..83N}. The short GRB 170817A was the first electromagnetic counterpart coincident with the gravitational wave event GW 170817A which unambiguously confirmed that BNS mergers are at least part of the mechanism that produces SGRBs~\citep{2017ApJ...848L..12A}. Furthermore, in some events, it was found that the magnetar giant flare and underlying supernova components are also related to SGRBs which indicates that this mysterious phenomenon has multi-faceted characteristics intrinsically~\citep{2021Natur.600..621C}. Since GRBs were first detected by the Vela satellite in 1967~\citep{1973ApJ...182L..85K}, more than half a century has passed since the GRB field is still the frontier of multi-messenger astronomy. With generations of instruments involved in this research field, the method of GRBs' localization is described below. The space detectors in the high-energy band, such as \textit{Swift} and \textit{Fermi} can localise the burst position with GRBs' prompt emission and then circulate this result to the astronomical community all over the world through the Gamma-ray Coordinates Network (GCN). Ground facilities can follow up their position in order to capture their afterglow. Their optical counterparts play an important role in providing key information about the bursts themselves, such as color index and redshift, as well as the physical properties of their circumburst medium and host galaxies. However, observing the afterglows of GRBs remains challenging due to their short temporal scale, fast-decreasing brightness and faint afterglows. To address this, the BOOTES framework has implemented a strategy for GRB follow-up observations that aims to minimize the time delay between receiving the position and starting the observation, based on its autonomous reaction to triggers. Once the trigger is received by the telescope server, the narrow/wide-field telescopes slew to the burst position automatically as soon as a couple of seconds if the position is reachable. Meanwhile the all-sky camera will keep monitoring the sky in case that the burst has a very bright counterpart in the optical.

\begin{figure}[!h]
	%\centering
	\includegraphics[scale=0.5]{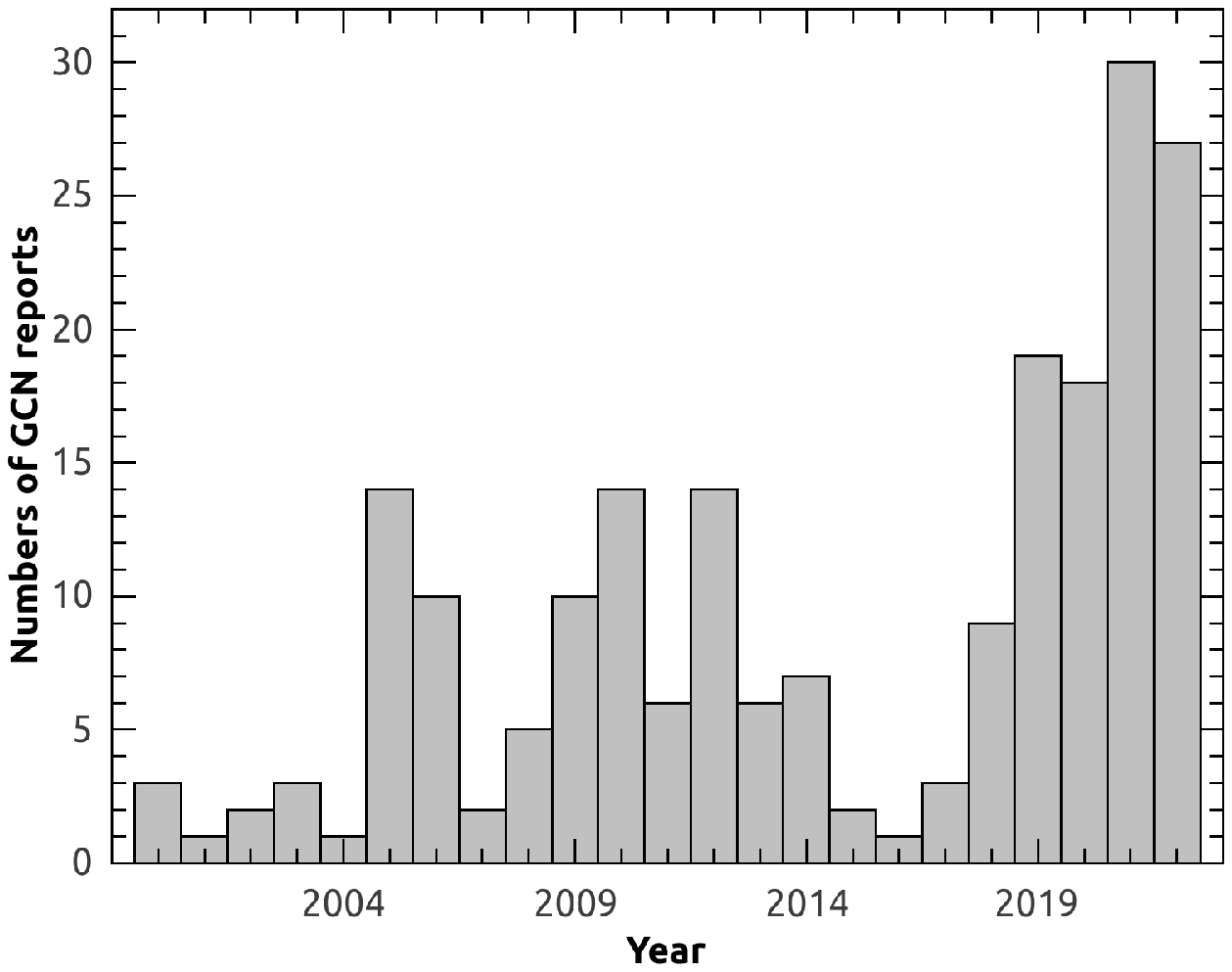}
	\includegraphics[scale=0.18]{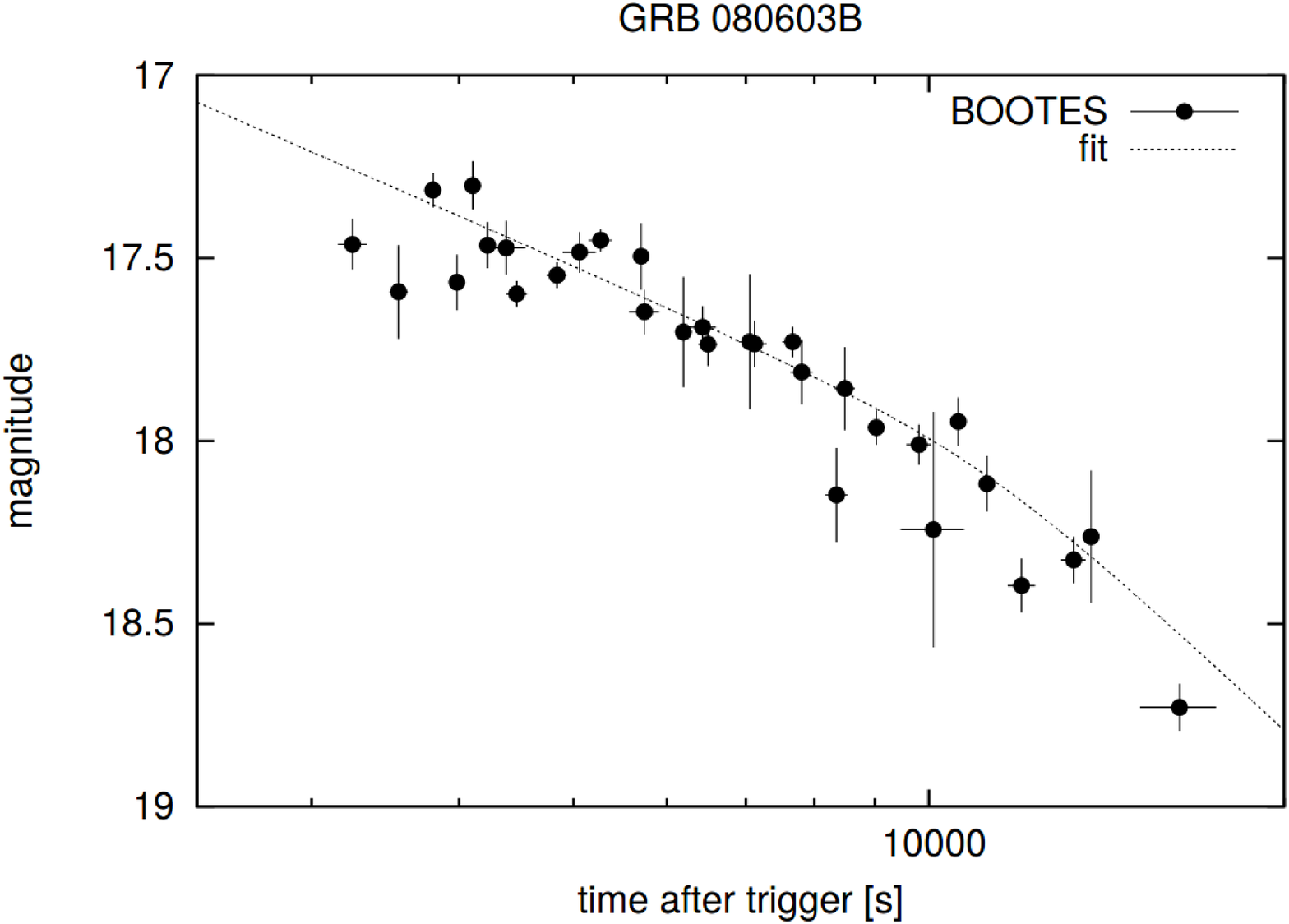}
	\caption{Left panel: BOOTES network GCN reports statistics until 2022. Right panel: GRB080603B optical observations with the BOOTES network with a smooth broken power-law that can fit well the afterglow~\citep{2012AcPol..52a..34J}.}
	\label{fig:BOOTESGRBobservation}
\end{figure}

Since the first BOOTES network operations, one of the main goals has been to observe all triggered bursts in order to detect their optical afterglows. However, their intrinsic faintness limited their detection rate. Totally, there have been 196 reports published on GCN Circulars for early time observations of 182 GRBs based on BOOTES results \citep{2010AdAst2010E..85J, 2016AdAst2016E..12J}. For 48 of these events we have found/confirmed their optical afterglows (see Appendix: Table~\ref{tab:BOOTESGRBs}). Other observations have not reported in this manner due to the timeliness of their observations but are published elsewhere. The count plot of the circular number (see in Figure~\ref{fig:BOOTESGRBobservation} left panel) clearly shows that the early detection rate has increased with the completeness of the BOOTES network construction and the following upgrades. The automation makes it possible to reach the burst position in 3 s after trigger and observations can provide detections as deep as 21 mag. In some cases, follow-up observations are executed by several telescopes from different sites. Though the all-sky camera only provides an upper limit of $\sim$10 mag, the observations can be performed simultaneously/semi-simultaneously during the prompt emission phase which can be used to detect the early phase of the event similar to the naked-eye burst, e.g. GRB080319B \citep{2008Natur.455..183R}. This proves the capability of the BOOTES network to search for and provide the early observations of the optical afterglows of GRBs. These continuing observations can be used to constrain the afterglow evolution models. For example, in the case GRB080603B (see Figure~\ref{fig:BOOTESGRBobservation} right panel), the optical light curve can be fitted with a broken power-law showing a smooth transition between two decay epochs from $\alpha_{1}=-0.55\pm0.16$ to $\alpha_{2}=-1.23\pm0.22$~\citep{2012AcPol..52a..34J}. Thanks to the spectral indices measured, it can be suggested that this burst is a case of a stellar wind profile expansion in a slow cooling regime. 

\subsection{Fast-radio bursts}
Fast radio bursts (FRBs) are a new type of high energy transients discovered in the 21st century \citep{2007Sci...318..777L}. They were first named "Lorimer bursts" because of his contribution to the first FRB event detection with the Parkes telescope in Australia in 2007. This phenomenon has a characteristic time-scale of a millisecond duration in its MHz-GHz radio emission with a high dispersion measure value \citep{2013Sci...341...53T}. With the accumulation of new detections, it has been found an isotropic sky distribution instead of the high latitude region distribution which suggests a cosmological origin. Since their first report, a number of radio facilities have been conducted to search for FRBs, such as the Canadian Hydrogen Intensity Mapping Experiment (CHIME), the Deep Synoptic Array (DSA) and the Five-hundred-meter Aperture Spherical radio Telescope (FAST). With their joint efforts, the FRBs detection rate has increased and it has been found that there are two main types of bursts, i.e. repeating/non-repeating, among which most of them are the non-repeating cases. Since their millisecond duration, it is difficult to take follow-up observations except for the repeating ones. The first repeating event, FRB121102A, was identified with the Arecibo Observatory and it was found to have a 157-day cycle which enables the precise localization of the burst place \citep{2014ApJ...790..101S}. It has been found that the host galaxy is a low-metallicity star-forming dwarf galaxy at a redshift of z=0.193. Recently, CHIME published its new catalogue \citep{2021ApJS..257...59C} on FRBs hunting which included 535 events and 61 bursts from 18 recorded positions while other 474 events are one-off bursts. With these repeating FRBs, observations in other wavelengths become possible, spanning from optical, X-ray to gamma-ray, in order to place constraints on their radiation mechanism. There are several ways to search for the optical or high-energy counterparts of FRBs, such as using the same procedure as for triggering a GRB observation, monitoring burst fields with wide field telescopes, or targeting the repeating FRBs directly. Through these efforts, we are learning more about bursts, their multi-band counterparts, and the host galaxies they belong to, which is helping us to better understand this mysterious phenomenon.

\begin{figure}[!h]
	\centering
	\includegraphics[scale=0.2]{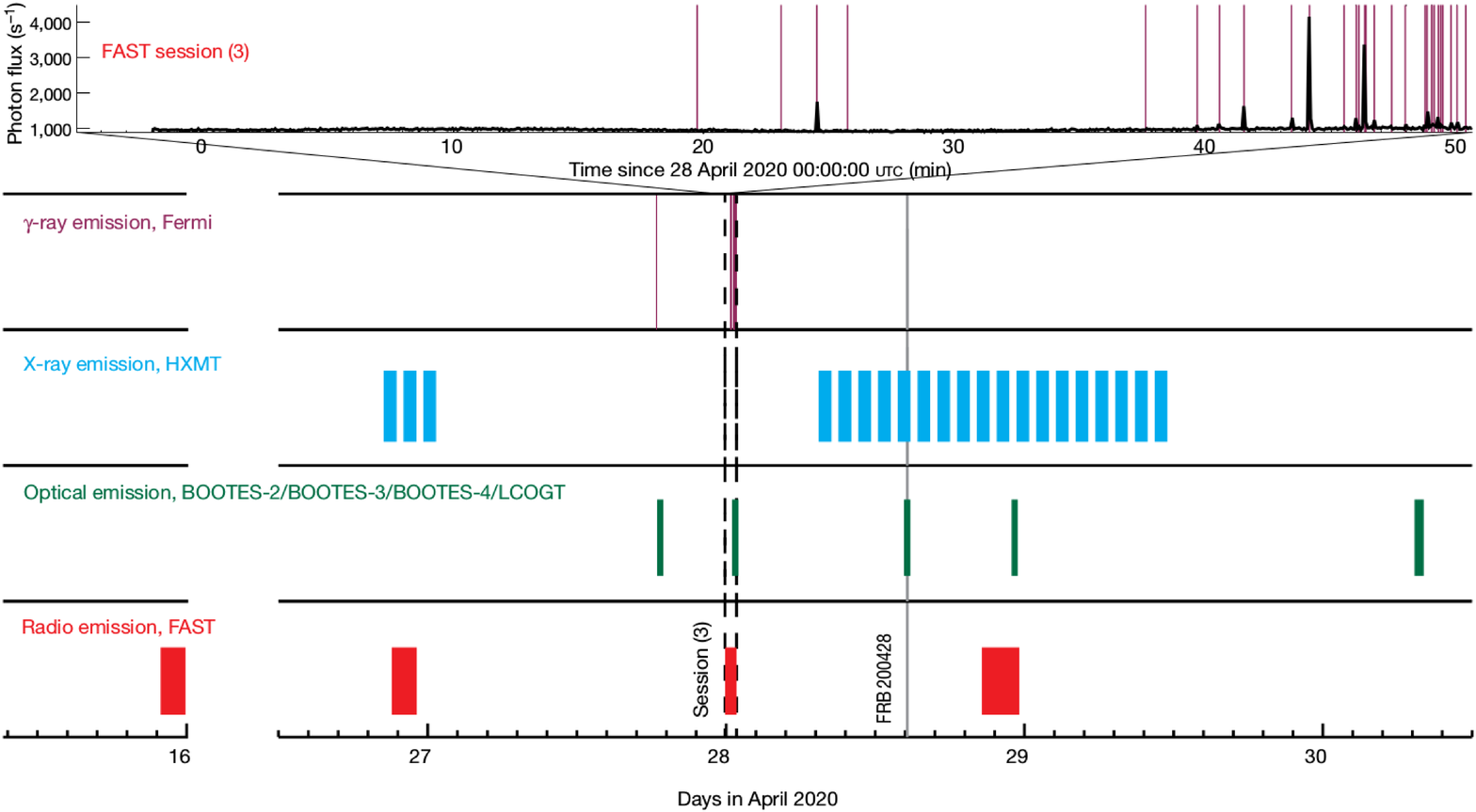}
	\caption{The timeline of SGR 1935+2154 observations with Fermi, HXMT, BOOTES, LCOGT and FAST, i.e. from radio to gamma-ray bands \citep{2020Natur.587...63L}.}
	\label{fig:FRB200428}
\end{figure}

As the BOOTES network is composed of several narrow and wide-field cameras \citep{2012ASInC...7..313C}, this has proven to be a good platform for optical FRB counterpart searching. Like for the GRB follow-up observations mode, observing campaigns in other wavelengths can trigger the BOOTES telescopes to react simultaneously/semi-simultaneously for obtaining images in the burst active phase. On April 28 2020, a bright FRB from the Galactic magnetar SGR 1935+2154 was captured with the CHIME radio telescope and STARE2 (Survey for Transient Astronomical Radio Emission 2) radio array \citep{2020Natur.587...54C,2020Natur.587...59B}. Meanwhile, the FAST telescope observed the same position in four sessions but a burst happened between the third and fourth sessions \citep{2020Natur.587...63L}. During the third 1-hr session, the magnetar became very active and emitted 29 bursts as observed by the \textit{Fermi} satellite. Following the previous high-energy detection trigger of this magnetar, the BOOTES network responded it globally and there were simultaneous images taken from the BOOTES-3 station during its emitting episode. See Figure~\ref{fig:FRB200428} for the timeline of the multi-wavelength observations. The multi-burst phase was observed in gamma-ray, optical and radio. In a series of images in the Z-band obtained during the burst and in the simultaneous 60 s exposure frames we got a limiting magnitude of 17.9 mag. Considering the extinction correction, this limit corresponds to a peak flux of $F_{\mu,opt}\leq$4.4kJy for 1 ms combined with a radio counterpart flux of $F_{\mu,FRB}\geq$1.5MJy for 1 ms that lead to the flux ratio between fast optical bursts (FOB) and fast radio bursts should follow $\tau\leq10^{-3}$. This stringent limit gives the first meaningful constraint on the FOB model parameters. Similarly, other facilities also follow up FRBs, such as the Zwicky Transient Facility, \citep[ZTF][]{2020ApJ...896L...2A}, the Ground-based Wide Angle Camera \citep[GWAC,][]{2021ApJ...922...78X} and Apache Point Observatory \citep[APO,][]{2021ApJ...907L...3K}. However, the non-detection of these elusive optical counterparts presents a challenge for the BOOTES network in the future.

\subsection{Gravitational waves electromagnetic counterparts}
Massive stars are the most interesting and mysterious objects in the sky since their late evolution could be the best laboratory to produce catastrophic phenomena during the birth of black holes and neutron stars and due to their interaction with surrounding objects \citep{2009ARA&A..47..107H}. Therefore, they constitute an ideal natural environment to search for new physics. Theoretically, most of the energy in the two compact objects' merger process is released through the electromagnetic radiation that is produced together with the gravitational waves (GW) propagation containing part of the potential energy. This could be a new probe to investigate the information about their host and the characteristics of the objects (e.g. mass, spin-orbit, etc.). The first GW detector was proposed and designed in the 1960s with large cylinders of aluminium \citep{1968PhRvL..20.1307W}. After about half a century of research, the new generation of GW detectors used the laser interferometry method and there are already several main detectors built, such as the Laser Interferometer Gravitational-wave Observatory (LIGO, Livingston and Hanford at America) and Virgo interferometer (Virgo, Pisa at Italy). More detectors are under construction to join observation in the near future, such as the Indian Initiative in Gravitational-wave Observations (IndIGO). During the LIGO-Virgo scientific operation, the first GW event was detected on September 14, 2015, which was confirmed to be generated from two $\sim$30 solar mass black holes merging at a distance of $410^{+160}_{-180}$Mpc \citep{2016ApJ...826L..13A}. Together with the GW signal, the gravitational potential released in the form of electromagnetic (EM) radiation, reached Earth simultaneously which could be used to constrain their related physical processes. Optical observations, which have well-developed techniques in the visible wavelength range, can be particularly useful in searching for the EM counterpart of GW events. Furthermore, the optical follow-up observation of the optical counterparts can provide multi-colour evolution to distinguish between different physical models. Along with the observations at other wavelengths, this will push multi-messenger astronomy ahead. Currently, the first discovery of an EM counterpart of a GW signal was found to be associated with a short gamma-ray burst, i.e. GRB170817 \citep{2017ApJ...848L..12A}. Due to the transient's short time scale, a telescope with fast localization and rapid follow-up capabilities was key for studying its properties. Since the BOOTES network is a robotic telescope system, it is well-suited for studying this type of transient event.

\begin{figure}[!h]
	\centering
	\includegraphics[scale=0.2]{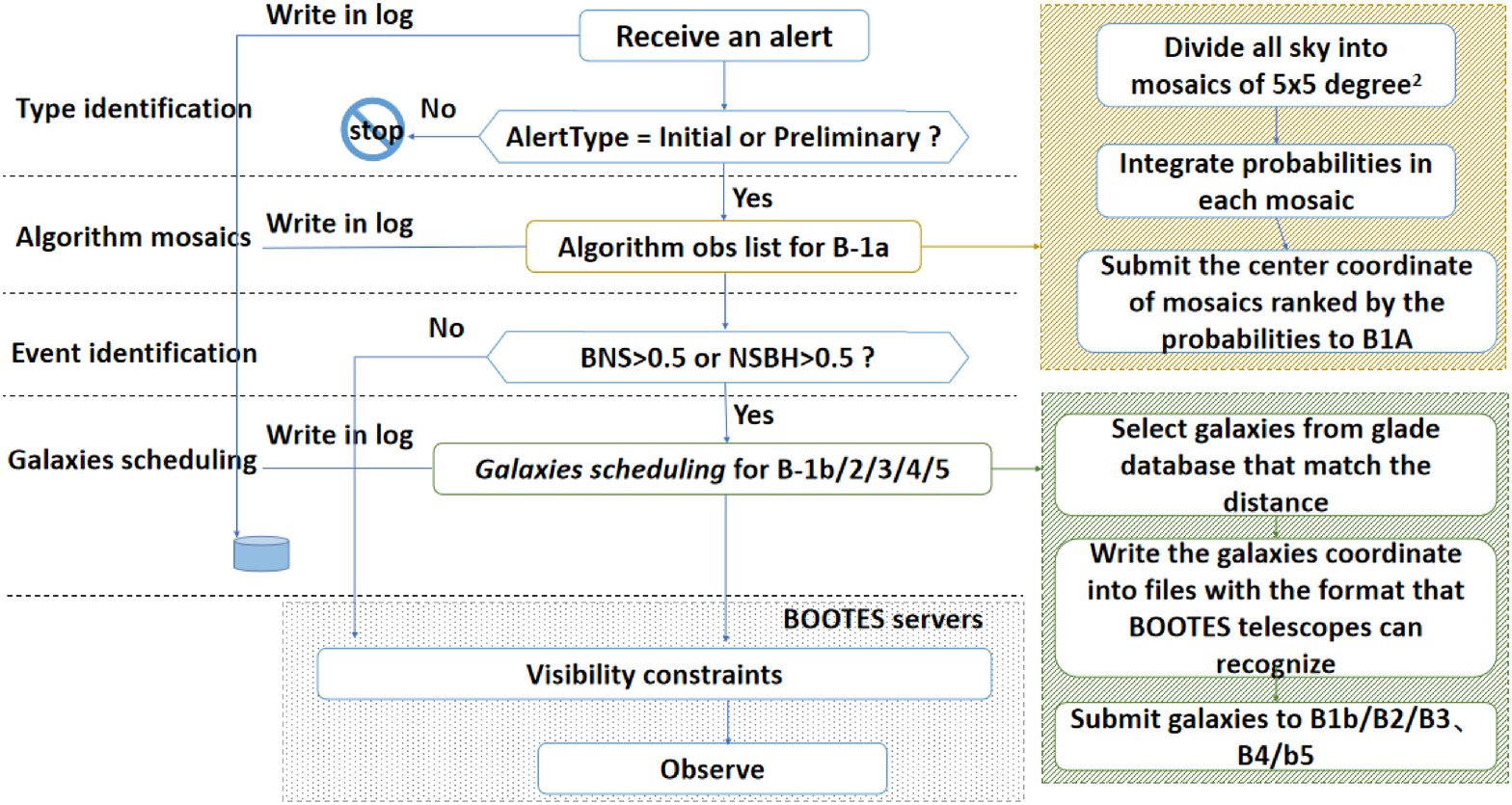}
	\caption{The strategy used in the BOOTES network for the GW observation.}
	\label{fig:GWstrategy}
\end{figure}

Since the LIGO-Virgo completed their construction, they have already made three joint scientific observing runs: September 12, 2015 - January 19, 2016 (O1), November 30, 2016 - August 25, 2017 (O2) and April 1, 2019 - March 27, 2020 (O3) \citep{2021arXiv211103606T}. The localization uncertainty was of several thousands $deg^{2}$ until Virgo's joint later which decreased the error region down to 28 $deg^{2}$ in the case of GW170817 \citep{2017ApJ...848L..12A}. Facing bigger error regions, the strategy used in the BOOTES network is shown in Figure~\ref{fig:GWstrategy}. The BOOTES Network has both wide and narrow FOV telescopes, which can observe the new events in different ways. When a new GW alert is received, the whole region is observed using a mosaic approach with the wide-field camera B1A. If that alert is identified as being related to neutron stars (NS), then the galaxy candidates in the error region are given higher priority follow-up observation using the narrower FOV telescopes.

During O1, the BOOTES network observed the event GW150914 with the CASANDRA camera at the BOOTES-3 station~\citep{2016ApJ...826L..13A}. 
The image was taken simultaneously and no prompt optical counterpart was found. Due to the poor weather conditions, it only provided an upper limit of 5 mag. While this value sets a wide constraint, the BOOTES observation provided the earliest image corresponding to the first GW alert and demonstrated that its wide field coverage is sufficient for searching for GW counterparts. During O2, the milestone event of GW170817 was detected in images taken from the BOOTES-5 telescope. The magnitude recorded was $r=18.2\pm0.45$ which is brighter than the predicted flux of optical afterglows but is consistent with other contemporaneous measurements~\citep{2018NatCo...9..447Z}. During O3, there have been a total of 72 alerts, with 16 of them being NS-related merger events. The BOOTES network followed up on 55 of these events, including 13 NS-related mergers. Although the 76\% of these alerts triggered BOOTES telescopes, there was not any new object detected in these images thus giving only a typical 3-$\sigma$ upper limit of 20 mag ~\citep{2021RMxAC..53...75H}.

\subsection{Neutrino burst and blazar monitoring}
The detection of neutrinos is another important aspect of multi-messenger astronomy which contains pieces of information essential for the understanding of high-energy events since these elementary electrically neutral particles have been detected arising from some nearby astrophysical sources \citep{2000ARNPS..50..679L}. They only interact through the weak force which makes them very difficult to be detected. On the other hand, they can point out the source position directly. Tracking the Cherenkov radiation photons is the way to capture these events nowadays. The IceCube observatory in Antarctica is the largest neutrino detector currently. It consists of 86 strings, each connecting 60 digital optical modules, which are distributed in the Antarctic ice within one square kilometre area evenly, i.e. a km$^{3}$ cubic of ice as its volume providing the detectable energy range from 0.1 TeV to beyond 1 EeV \citep{2013PhRvL.111b1103A}. Similar ideas to design the neutrino detectors are also used such as ANTARES which is located in the Mediterranean Sea. For the time being, neutrinos from SN1987A in the Large Magellanic Cloud were detected by the Kamiokande experiment which was the first astrophysical source of neutrinos ever identified \citep{1987Natur.330..142S}. The second detection of astrophysical origin was in the direction of the flaring active galactic nucleus AGN TXS 0506+056 \citep{2018Sci...361..147I}, a BL Lac object. Although several models have predicted the association between neutrino and other catastrophic events in the Universe, no direct coincident evidence supporting this point has been made so far.

The BOOTES network also responds to neutrino events as one of its scientific targets and uses a similar method to the one used for GRB follow-up observations. Once an alert of this kind of event is received, the BOOTES telescopes point towards their origin directly to search for any new candidates that may be related to the neutrino source. For example, the second neutrino event detection, i.e. IceCube-201114A, was found to be related to the blazar NVSS J065844+063711 which was included in the \textit{Fermi}-LAT fourth source catalogue with the name of 4FGL J0658.6+0636. The BOOTES-2 station monitored it for several nights in order to check its short-time variability \citep{2022icrc.confE.955D}. Besides, long-term optical monitoring of variability in blazars was attempted for the object S5 0716+714 \citep{2019Ap&SS.364...83W} which is a candidate of high energy neutrino and high energy radiation. The BOOTES-4 station continued monitoring this source in multiple bands for several years and was the first facility to discovery its long-term variation pattern: a strong flatter when brighter (FWB) trend at a low flux state and then a weak FWB trend at a higher flux state, which was interpreted as the acceleration and cooling mechanisms of different electron's populations in the relativistic jet.

\subsection{Outreach}
In addition to the scientific results mentiond above, the BOOTES network also plays an important role in the public science education. By increasing scientific literacy and opportunities among young people, it helps to foster scientific vocations and provides opportunities for interested individuals to get involved in real research projects.

\subsubsection{GLORIA}
In 2009, the use of a worldwide network of robotic telescopes for educational purposes was proposed to the European Union, dubbed the GLObal Robotic telescopes Intelligent Array~\citep[GLORIA,][]{2013EAS....61..483M}. This intelligent array devoted a fraction of the available observing time of existing telescopes for public use and it was the first large-scale robotic telescope network with free access. This project was initially funded in October 2011 under the support of the European Union within the Seventh Framework programme (EU FP-7) for research and technological development including demonstration activities and lasted three years. Finally, twelve institutions from seven countries (Spain, Chile, Czech Republic, Ireland, Italy, Poland and Russia) participated in this project and made use of 18 telescopes working in different scientific fields. As part of the Spanish contribution, three of the BOOTES Network telescopes (BOOTES-1, BOOTES-2, and BOOTES-3) were part of GLORIA. By using web 2.0 technology, anyone could freely access and obtain nearly real-time images of the night sky using a working telescope. There were two modes of observational application: the online mode (sun observation, interactive night sky observation, scheduled night sky observation) and the offline mode (archival images in the database of GLORIA or from other databases including the European Virtual Observatory). The main goals of this project were to increase the number of telescopes and the number of scientists and citizens interested in astronomy, in order to expand the scope of research in these areas. By making knowledge freely accessible through this project, public motivation to engage in science education was enhanced and barriers to access were reduced. During the project, 5 astronomical events (4 eclipses and a transit of Venus) were broadcasted through the GLORIA network with associated educational activities in schools of the partner countries. Furthermore, scientific research was also conducted using GLORIA, including the observation of the active star DG CVn \citep{2015MNRAS.452.4195C} and the eccentric eclipsing binaries in our galaxy \citep{2018A&A...619A..85Z}. 
\subsubsection{ScienceIES}
Facing the fact of the decreasing number of students in Science, Technology, Engineering and Mathematics at Spanish Universities in the early 2000s, an educational project at the Andalusian level started to bring high school students to work together with scientists in Andalusia (the southernmost autonomous community in Peninsular Spain) in 2010. This project named "Proyecto de Iniciación a la Investigación e Innovación en Secundaria en Andalucía (PIIISA)" and was then dubbed with ScienceIES \citep{2021RMxAC..53..151C}. It is a distinctive education way to teach natural sciences including Astronomy to high-school students (at age of 15-17). This project provides the chance to join together high school students, their teachers and scientists to solve a practical problem and undertake the experiment in schools or laboratories. Students involved in these proposed projects are required to write evaluations, take interviews and complete surveys under the supervision of scientists. Normally, during the middle of every academic year, introductory lectures for ScienceIES are held at university laboratories or research centers, spaced 3 days apart. These experiments and/or observations needed in this project are carried out at the same place to obtain the preliminary result which are subsequently evaluated by their high-school teacher and supervisor after being presented in public. Finally, research projects from each province in Andalusia that receive high ranking scores will be presented in the form of oral talks or posters at a one-day conference (held in May) in the main town at the regional level. The BOOTES network has been donating a portion of its telescope observing time to various astronomical research projects in PIIISA since 2013. Previous projects have included investigations of X-ray binaries, the local group of galaxies, and meteor storms.

\section{Robotic telescope networks worldwide}
Robotic telescopes have demonstrated many advantages over conventional telescopes, including fast reaction times, long-term monitoring capability, unmanned operations, lower costs, and more. It is clear that a worldwide network of robotic telescopes, comprising multiple sites, will enhance these advantages and improve observational efficiency. Like the BOOTES Network of Robotic Telescopes, there are several projects that involve deploying a network of robotic telescopes. Some of these projects have already been decommissioned. The following is a list of these projects.

\begin{itemize}
\item ROTSE-III (Robotic Optical Transient Search Experiment)\footnote{http://www.rotse.net/}: A network which consisted of four 0.45m diameter robotic Cassegrain telescopes distributed in different counties. Since 1998, ROTSE was devoted to searching optical transients and it operates with a wide FOV (1.85$^{\circ}\times$1.85$^{\circ}$) camera onto f=1.9 telescopes without filters \citep{2003PASP..115..132A}. 
\item LCOGT (Las Cumbres Observatory Global Telescope)\footnote{https://lco.global/}: It is a network composed of 25 telescopes, including 2$\times$2-m, 13$\times$1-m and 10$\times$0.4-m, where the 2-m telescopes were operated by RoboNet \citep{2009AN....330....4T} and purchased by LCO. They are distributed in 7 sites worldwide with the Ali Observatory station (the only one in Asia, still under construction in 2023). Optical imagers and spectrographs are designed and used in this network. It is operated by using a software scheduler which continuously optimizes the observing schedule of each telescope to monitor any target in the night sky \citep{2014SPIE.9149E..12P}.

\item MASTER (Mobile Astronomical System of Telescope-Robots)\footnote{http://observ.pereplet.ru/}: This network has 9 sites worldwide ( just missing an station in Oceania to achieve complete coverage ) which all are installed with a two-tube aperture system. Each tube is a 0.4-m telescope equipped with a 4k $\times$ 4k CCD camera with a scale of 1.85"/pixel and a universal photometer with Johnson-Cousins (BVRI) and polarizer filters. Its main scientific goals are not limited by the prompt optical emission of GRBs but the discovery of uncataloged objects \citep{2013ARep...57..233G}.
\item SONG (Stellar Observations Network Group)\footnote{https://phys.au.dk/song}: It is a Danish-led project to design and construct a global network of 1-m class robotic telescopes in 8 nodes to undertake long-time monitoring. So far 3 nodes have been built at Teide Observatory, Mt. Kent Observatory and Delingha Observatory. Lucky-imaging camera and a high-resolution spectrograph have been mounted in these telescopes to suit the requirement of the main observational goals \citep[asteroseismology and gravitational microlensing; ][]{2009ASPC..416..579G, 2016RMxAC..48...54A}. 
\item TRTN (Thai Robotic Telescope Network)\footnote{https://trt.narit.or.th/}: A network of several 0.6m and 0.7m telescopes at 5 sites worldwide which is used for astronomical research, education and public outreach activities. Both telescopes have a set of Johnson-Cousins (BVRI) filters and a 2k$\times$2k CCD camera to provide photometric observation \citep{Soonthornthum2017}.

\item TAROT (Télescopes à Action Rapide pour les Objets Transitoires)\footnote{https://web.tarotnet.org/}: It is a French-led network which consist of 2$\times$0.25-m (FOV 1.8$^{\circ}\times$1.8$^{\circ}$), 1$\times$0.18-m (4$^{\circ}\times$4$^{\circ}$) telescopes at 3 sites worldwide in addition to the 1-m Zadko telescope (FOV 20"$\times$20") at Gingin Observatory. They can be used to take photometric images with Sloan filters for multiple science goals, such as resident space objects, GRBs \citep{Boer2017TAROTAN}.
\item TAT (Taiwan Automated Telescope) Network: This network is planned to install several robotic telescopes worldwide. Each site has installed a 9 cm Questar telescope to provide a variable FOV but it has been set to 0.62 degrees square currently. Images in UBVRI filters can be obtained which are dedicated to photometric measurements of stellar pulsations to study the stellar structure and evolution \citep{2010AdAst2010E..13C}.
\item MicroObservatory: It is a worldwide network that comprises five robotic 0.91 m tall, 15 cm aperture, reflecting telescopes. Each unit is equipped with photometric filters (BVRI) and one neutral density filter which can provide a FOV of $\sim$1 square-degree. This project was dedicated to public education but also can be used in the scientific field, such as transiting exoplanets monitor \citep{2020arXiv200713381F}.
\item SPECULOOS (Search for habitable Planets EClipsing ULtra-cOOl Stars): This is a network of 1m-class robotic telescopes searching for transiting terrestrial planets around the nearest and brightest ultra-cool dwarfs. It has two main nodes, one in each hemisphere. The southern one consists of four 1-m telescopes and the northern one has the same plan but just has 1 telescope installed so far. A thermoelectrically-cooled camera with a near-IR-optimized deeply depleted 2k$\times$2k CCD detector, which is sensitive from ~350 nm (near-UV) to ~950 nm (near-IR), has been installed in every site to cover 12'$\times$12' sky. Currently, Sloan-g'r'i'z' filters, a special exoplanet filter 'I+z' and a blue-blocking filter are installed in the filter wheel in order to observe redder objects, such as mid-to-late M dwarfs \citep{2018SPIE10700E..1ID, 2020MNRAS.495.2446M}.
\item ARTN (Arizona Robotic Telescope Network): This network aims to upgrade and integrate multiple small to medium size telescopes (1-3 meters) to have multiple capabilities, including photometry, spectroscopy, wide-field optical imaging, rapid response and monitoring and making the sky survey. Nowadays, Steward Observatory's 1.55 m Kuiper telescope and Vatican Advanced Technology Telescope are in the first phase of the upgrade \citep{2018SPIE10704E..2HW}
\item Skynet\footnote{https://skynet.unc.edu/}: It is a robotic telescope network spanning four continents which can schedule targets through the control web interface. At present, it has not only optical telescopes whose size ranges from 35 to 100 cm but also a radio telescope, the Green Bank Observatory's 20-m diameter radio telescope. Hence, the multi-color optical image and the timing/mapping observations in the radio band are available to support professional astronomers and public education use in various astronomy research fields \citep{2019ApJS..240...12M, 2022arXiv220209257R}.
\item SMARTNet (the Small Aperture Robotic Telescope Network): It was planned to initiate a robotic telescope network worldwide with a minimum of 6 stations and there are 3 stations finished construction already. Regarding its multiple objectives, a two-telescope setup with different apertures is favoured which includes a 20 cm (FOV 3.5$^{\circ}\times$3.5$^{\circ}$) plus a 50 cm (FOV 36'$\times$36') telescopes to provide a deep survey of faint objects and a fast survey of bright objects, respectively. Later, the wide field telescope upgrade to a 25 cm diameter telescope (FOV 2.3$^{\circ}\times$2.3$^{\circ}$). The obtained optical images will are used to survey the Geostationary Orbits (GEO) objects and label them \citep{dlr123984,dlr146889}.
\item OWL-Net (Optical Wide-field patroL Network): This network is composed of 0.5-m wide field optical telescopes operated in a robotic manner and distributed in 6 sites in the northern hemisphere. Each telescope has installed a 4k$\times$4k CCD camera with Johnson (BVRI) filters and a FOV of  1.1$^{\circ}\times$1.1$^{\circ}$. Although its main objective is to monitor Korean Low Earth Orbits (LEO) and GEO satellites to maintain their orbital information, the astronomical mode is also accessible for photometric images \citep{2018AdSpR..62..152P}.
\end{itemize}
In addition to these worldwide robotic telescopes networks, there are also several single-site robotic telescopes networks, such as the Ground Wide Angle Camera Network \citep[GWAC; ][]{2021PASP..133f5001H}, and the MINiature Exoplanet Radial Velocity Array \citep[MINERVA; ][]{2015JATIS...1b7002S}. Currently, the 2-m class robotic telescope is the biggest size of robotic telescopes, such as the Faulkes Telescopes owned by LCOGT, and most telescopes in these networks are of small size. From a construction perspective, the majority of these robotic networks were newly designed and built. However, the upgrade of existing telescopes is also an option, such as with the ARTN. In terms of research objectives, both multi-objective networks, such as LCOGT and Skynet, and single-objective networks, such as TAT and SMARTNet, have been established. Regarding time domain astronomy, an alert system is necessary for the observation of transients observation by providing a Target of Opportunity (ToO) on robotic telescopes, such as MASTER and TAROT. From the perspective of the observation mode, all of these networks have imaging capabilities, some with multi-colour capability or even polarization mode, such as MASTER, but fewer have spectroscopic observation mode, such as LCOGT and SONG. Near-IR observations are even fewer in number, only SPECULOOS. From the user perspective, some networks are used exclusively for scientific research, such as OWL-Net, but some networks are also involved with public education programs, such as MicroObservatory and Skynet. While most of these robotic systems are still expanding their deployment to additional stations. 

Among them, the BOOTES network is a Spanish-lead network of medium-size (0.6m) small aperture robotic telescopes with the following capabilities: 
\begin{itemize}
    \item multi-band imaging and optical spectroscopy,% (both imaging from visible to near-IR and optical spectroscopy),
    \item multi-objective (GRBs, FRBs, GWs etc), 
    \item ToO reactive (with long-period monitoring observation capability),
    \item for multiple purposes (both scientific and public science education) observations. 
\end{itemize}
The BOOTES Network is composed of both wide and narrow field cameras in each individual astronomical station. 

Regarding the transient automatic follow-up observations, the MASTER network shares several overlapping scientific objectives with the BOOTES project. They both can quickly react to new events' triggers and have their own unique geographical locations (with no overlap between them). With bigger diameter telescopes apertures, the BOOTES network can provide deeper limiting magnitudes whilst the tube design of the MASTER network makes it suitable for polarization measurements. In addition, the BOOTES network can provide low-resolution spectroscopy for bright objects. Currently, the MASTER network has two observatories in the southern hemisphere. With the new station BOOTES-7 deployed in 2022, the BOOTES Network has became the first Global Network of Robotic Telescopes present on all continents.

\section{Conclusions and Future prospects}
The Burst Optical Observer and Transient Exploring System (BOOTES) is the first Global Network of Robotic Telescopes present on all continents. Nowadays, it is composed of 7 stations already installed in 6 countries, i.e. Spain, New Zealand, China, Mexico, South Africa and Chile. This network makes use of both wide-field and narrow-field cameras. An ultra-lightweight telescope concept proposed for the BOOTES Network has allotted to deploy 6 copies in the above-mentioned countries which provide the capability of fast slewing. With the complementary all-sky camera, the BOOTES network can monitor any night sky in both the northern and southern hemispheres. The BOOTES telescopes are equipped with weather information detectors and a self-designed control system based on the ASCOM platform. This allows the telescopes to automatically receive alerts and schedule observations in a matter of seconds. It is also possible to conduct remote observations through HTTP communication.

Taking full advantage of the BOOTES Global Network of Robotic Telescopes, its scientific goals are set to quickly react to high-energy transients alerts (GRBs, FRBs, GWs , etc.) and also to support scientific space missions. In regards to the BOOTES Network's observations of astronomical transients, each BOOTES telescope automatically reacts to incoming triggers by slewing to the source position using different strategies depending on the object's type. The BOOTES telescopes can react as fast as in 8 s to start gathering the observations (down to 20 mag, such as in the case of GRBs observations). With the gradual completion of the network, the detection rate of transients has increased. The rapid follow-up observations from early to late-times executed by BOOTES have been used to constrain the GRB's models. In case of FRBs, for the time being, there is still no clear optical detection related to FRBs themselves. For neutrino events, the upper limits obtained so far can set meaningful constrains on the existing models. Electromagnetic counterpart searches for GWs in the past GW detectors scientific runs on BOOTES network were executed perfectly. The optical counterpart to the milestone event, dubbed GW170817, was detected by BOOTES-5 (the only Spanish installation doing so). Even though there has not been any electromagnetic counterpart related to a GW detected in the O3, 76\% of triggers have been followed up, providing a typical upper limit of 20 mag which can be used to constrain model parameters. 

On top of this, the BOOTES network has contributed to public outreach through the EU FP-7 funded GLORIA project, which allows citizens to view the night sky for free through web 2.0. It also allows astronomers to gather additional scientific data. Currently, astronomy projects at the high-school level, such as ScienceIES in Spain, are giving teenagers the opportunity to experience real scientific research and obtain scientific results under the guidance of professional researchers.

The complete BOOTES network will make the transients' follow-up more efficient. The increasing number of follow-up observations regarding GRBs, FRBs, neutrino sources and the possible electromagnetic counterparts related to the fourth LIGO/Virgo scientific run (O4) in 2023 are expected to provide outstanding results. Robotic telescope networks are eager to contribute to these efforts and help shed light on the Cosmos.

\section*{Acknowledgments}
 YDH acknowledges support under the additional funding from the RYC2019-026465-I. MCG acknowledges support from the Ram\'on y Cajal Fellowship RYC2019-026465-I. AJCT acknowledges support from the Spanish Ministry project
PID2020-118491GB-I00 and Junta de Andalucia grant
P20 010168. MCG and AJCT acknowledge financial support from the State Agency for Research of the Spanish MCIU through the ``Center of Excellence Severo Ochoa" award to the Instituto de Astrofísica de Andalucía (SEV-2017-0709).
We also acknowledge to all colleagues of the scientific institutions hosting the BOOTES stations worldwide. Without them, the achievement of the BOOTES Network would have been much lower. Finally, the authors extend their thanks to the anonymous {\bf referees} for their valuable comments that helped to improve the manuscript substantially.

\bibliographystyle{frontiersinSCNS_ENG_HUMS} 
\bibliography{Robotic}

\end{document}